\documentclass[12pt]{article}
\usepackage{epsfig}

\parskip=\medskipamount 
plus.3em minus.5em \abovedisplayshortskip=.5em plus.2em minus.4em
\renewcommand{\thesection}{\arabic{section}}

\catcode`@=11

\@addtoreset{equation}{section} \@addtoreset{equation}{subsection}
\def\theequation{\ifnum\value{section}=0 \arabic{equation}\ignorespaces
\else \ifnum\value{section}=-1 A.\arabic{equation}\ignorespaces
\else \ifnum\value{subsection}=0
\thesection.\arabic{equation}\ignorespaces \else
\thesection.\arabic{subsection}.\arabic{equation}\ignorespaces
                             \fi
                        \fi
                   \fi}

{\catcode`\'=\active \def'{{}^\bgroup\prim@s}}

\catcode`@=12

\newcommand{\bq}{\begin{equation}}
\newcommand{\be}{\begin{equation}}
\newcommand{\fq}{\end{equation}}
\newcommand{\ee}{\end{equation}}
\newcommand{\bqr}{\begin{eqnarray}}
\newcommand{\beqs}{\begin{eqnarray}}
\newcommand{\fqr}{\end{eqnarray}}
\newcommand{\eeqs}{\end{eqnarray}}

\newcommand{\rf}[1]{(\ref{#1})}

\def\bop#1{\setbox0=\hbox{$#1M$}\mkern1.5mu
    \vbox{\hrule height0pt depth.04\ht0
    \hbox{\vrule width.04\ht0 height.9\ht0 \kern.9\ht0
    \vrule width.04\ht0}\hrule height.04\ht0}\mkern1.5mu}

\begin{document}
\thispagestyle{empty}

\vskip .6in
\begin{center}

{\Large\bf  Data Compression with Prime Numbers}

\vskip .6in

{\bf Gordon Chalmers}
\\[5mm]

{e-mail: chalmers@quartz.shango.com}

\vskip .5in minus .2in

{\bf Abstract}

\end{center}  

A compression algorithm is presented that uses the set 
of prime numbers.  Sequences of numbers are correlated 
with the prime numbers, and labeled with the integers.  
The algorithm can be iterated on data sets, generating 
factors of doubles on the compression.

\setcounter{page}{1}
\newpage
\setcounter{footnote}{0}

There are publicly available tens of million of prime 
numbers.  The bit complexity of these sets of numbers 
can be reduced exponentially by associating an integer 
count to the sets of these numbers.  Because there are 
an approximate $N/\ln(N)$ prime numbers below a number 
$N$, the complexity of the data set is reduced by an 
approximate $\ln(N)$.  

This reduction in the complexity of the prime number 
data set can be used to compress any set of integers 
as well.  By breaking the number up into bit sequences 
which are prime, and then labeling the numbers with the 
integers, a string of bits can be reduced in complexity 
by a $\ln(N)$ of the individual number sequences.  

For example, consider the number N=$101113$.  This 
number has the prime sequences of $101$ or $113$, 
which are the $26$ and $27$ the prime numbers.  
The two numbers could be registered by their indices, 
rather than the number sequences contained in $N$. 
The bit complexity is reduced to four digits, the 
$26$ and $27$, rather than the $6$ digits of $101113$.  
This reduction is not much for small numbers, but can 
be larger for prime number subsequences with ten or 
tens of digits.  The relevant ratio is the fraction 
of the number of prime numbers below a number $N$, which 
decreases as the number $N$ increases.  

The natural question is given a sequence of digits, 
what is the probability of finding a prime number 
contained in the subsequence.  Clearly, a number 
with only an even number of digits would fail this 
test, but real datasets arent expected to be composed 
of only even number digits.  

The chance of a random number $N$ of being prime is an 
approximate $1/\ln(N)$, which is the inverse of the 
number of digits.  Checking a subsequence of a 
number with $N_d$ digits requires summing the 
probabilities.  Checking the subsequences of a 
number with $P_d$ digits to $Q_d$ digits requires 
the sum, 

\bqr  
\sum_{P_d}^{Q_d} {1\over n} = \ln(Q_d/P_d)+C+{\cal O}(P_d/Q_d) \ ,   
\fqr  
which is always greater than unity if the numbers are chosen 
in a certain manner.  Clearly, if $Q_d$ and $P_d$ are chosed 
as a large ration, the probability will eventually be unity.    

The finite sums 

\bqr 
{1\over 5}+{1\over 6}+\ldots+ {1\over 9}\sim .75  \ ,  
\fqr  
\bqr 
{1\over 4}+{1\over 5}+\ldots+{1\over 9}>.99
\label{probabilities}
\fqr 
are almost unity.  The probability is unity 
to find a prime subsequence made of between $6$ to $13$ digits 
in a number $N$ containing more than $13$ digits.  There are 
public databases of the first $15$ million prime numbers, which 
consist of up to nine digits.  

The examples in \rf{probabilities} indicate that the public databases 
could be used to break up numbers into sequences of numbers which are 
prime.  The bit complexity in the reduction depends on the size of the 
sequences.  The bit complexity of a prime number of the size of $10^9$ 
is $30$ and that of its index with a number of the order $10^7$ is 
$23$; the ratio is a naive estimate of the 'worst' case scenarios of 
using sets of $23$ bits to label the prime number index versus the actual 
bit complexity of a nine digit number.  This ratio is $1.3$, which 
signals a $30$ percent compression factor.  An interesting aspect is that 
this algorithm can be iterated multiple times, with $30$ percent factor in 
each iteration (three iterations is a factor of $2.2$).   

A more efficient algorithm is to use two units.  The first states how 
many digits in the prime label, from $4$ to $7$, which has two bits.  
The second number specifies the prime number index.  The advantage of 
this is that not all prime numbers have $9$ digits (out of the exampled 
five to nine digits).  The index with four digits requires $13$ bits 
and the index with seven digits requires $23$ bits.  This should increase 
the compression factor to almost two, considering the distribution and 
probability of finding the prime number sequences.   (This version is 
similar to minimizing the bit vacancies in a byte or series of bits are 
not required to specify the 'color' of the data in certain compression 
schemes \cite{ChalmersData}.)

The examples listed pertain to prime numbers with up to $9$ digits, such 
as one billion.  Asymptotically the larger the number of digits in the 
prime number, the larger the compression factor will be.  Scanning for 
sequences with $P_d$ to $Q_d$ digits, with large numbers of digits is 
computationally intensive, but this will lead to larger compression factors.  
There is no bound to the compression factor given the distribution of 
primes $N/\ln(N)$; this says something about the entropy of the information.  

The iteration of the algorithm could easily produce compression factors 
of ten or so, given a front end for searching the prime sequences of the 
number $N$.  A database of the first 15 million primes would require a 
gigabyte of storage.   Also, this compression algorithm can be incorporated 
with existing algorithms.

\vfill\break

\end{document}